\documentclass[aps,prc,preprint,amsmath,amssymb,showpacs,preprintnumbers,superscriptaddress]{revtex4-1}
\usepackage{graphicx}
\usepackage{dcolumn}
\usepackage{bm}
\usepackage{color}
\usepackage{amsmath}
\allowdisplaybreaks[4]

\begin{document}

\title{Effects of rotation and valence nucleons in molecular ${\alpha}$-chain nuclei}

\author{D. D. Zhang}
\affiliation{State Key Laboratory of Nuclear Physics and Technology, School of Physics, Peking University, Beijing 100871, China}

\author{Z. X. Ren}
\affiliation{State Key Laboratory of Nuclear Physics and Technology, School of Physics, Peking University, Beijing 100871, China}

\author{P. W. Zhao}
\email{pwzhao@pku.edu.cn}
\affiliation{State Key Laboratory of Nuclear Physics and Technology, School of Physics, Peking University, Beijing 100871, China}

\author{D. Vretenar}
\email{vretenar@phy.hr}
\affiliation{Physics Department, Faculty of Science, University of Zagreb, 10000 Zagreb, Croatia}
\affiliation{State Key Laboratory of Nuclear Physics and Technology, School of Physics, Peking University, Beijing 100871, China}

\author{T. Nik\v si\'c}
\affiliation{Physics Department, Faculty of Science, University of Zagreb, 10000 Zagreb, Croatia}
\affiliation{State Key Laboratory of Nuclear Physics and Technology, School of Physics, Peking University, Beijing 100871, China}

\author{J. Meng}
\email{mengj@pku.edu.cn}
\affiliation{State Key Laboratory of Nuclear Physics and Technology, School of Physics, Peking University, Beijing 100871, China}

\date{\today}
\begin{abstract}
Effects of rotation and valence nucleons in molecular linear $\alpha$-chain nuclei are analyzed using a three-dimensional lattice cranking model based on covariant density functional theory. The structure of the mirror nuclei $^{16}$C and $^{16}$Ne is investigated as a function of rotational frequency. The valence nucleons, with respect to the 3$\alpha$ linear chain core of $^{12}$C, at low frequency occupy the $\pi$ molecular orbital. With increasing rotational frequency these nucleons transition from the $\pi$ orbital to the $\sigma$ molecular orbital, thus stabilizing the 3$\alpha$ linear chain structure.
It is predicted that the valence protons in $^{16}$Ne change occupation from the $\pi$ to the $\sigma$ molecular orbital at $\hbar\omega \approx 1.3$ MeV, a lower rotational frequency compared to $\hbar\omega \approx 1.7$ MeV for the valence neutrons in $^{16}$C. The same effects of valence protons are found in $^{20}$Mg, compared to the four valence neutrons in $^{20}$O. The model is also used to examine the effect of alignment of valence nucleons on the relative positions and size of the three $\alpha$-clusters in the mirror nuclei $^{16}$C and $^{16}$Ne.
\end{abstract}
\maketitle

\date{today}

\section{Introduction}\label{sec1}
Structure phenomena related to large deformations in exotic nuclei have been the subject of extensive experimental and theoretical studies. In heavy nuclei, a number of rotational bands have been observed that are based on axial quadrupole super-deformed and hyper-deformed states, characterized by the ratio 1:2 and 1:3, respectively, between the axis of the nuclear ellipsoid \cite{NyakPRL1984,TwinPRL1986,GalindoPRL1993,LaFossePRL1995,KrasznahorkayPRL1998}. In relatively light nuclei, not only pronounced deformations but also extremely elongated shapes such as, for instance, linear $\alpha$-cluster chain structure can occur.

The linear chain structure of three $\alpha$-clusters was firstly suggested in 1956~\cite{Morinaga1956}, and was used to describe the Hoyle state (the first excited $0^+$ state of $^{12}$C at $E_x=7.65$ MeV)~\cite{Hoyle1954} which plays a crucial role in the synthesis of carbon through the triple-$\alpha$ process, and it was observed in experiment soon after~\cite{Cook1957}. Much later, this state was also described as a gas-like structure~\cite{Fujiwara1980}, as well as  an $\alpha$ condensate-like state~\cite{Tohsaki2001,Suhara2014}. Since then, experimental and theoretical studies have been carried out in other $N=Z$ nuclei, such as $^{8}$Be~\cite{Datar2013,Garrido2013}, $^{16}$O~\cite{Chevallier1967,Suzuki1972,Flocard1984,Bender2003,Ichikawa2011,Yao2014,HePRL2014}, $^{24}$Mg~\cite{Iwata2015,Wuosmaa1992}. However, because of the antisymmetrization of single-nucleon wave-functions and the weak-coupling between $\alpha$-clusters, the linear chain configurations are difficult to stabilize. Therefore, to strengthen the stability of a linear chain structure, some additional mechanisms must be considered.

Additional stability of linear $\alpha$-chains can be obtained by the rotation of the nuclear system. At high angular momenta, the linear chain configuration with a large moment of inertia is favored because of the centrifugal force. However, very high angular momenta would also lead to a fission of the linear chain. The region of angular momentum in which the linear chain configuration is stabilized has been estimated theoretically for some $N=Z$ nuclei, for example, $^{16}$O~\cite{Ichikawa2011} and $^{24}$Mg~\cite{Iwata2015}. Another mechanism that can enhance the stability of linear chain configurations is the action of additional valence neutrons. For instance, when neutrons are added to
$N=Z$ nuclei and, especially when they occupy the $\sigma$-orbital (parallel to the symmetry axis of the linear chain), very elongated shapes are energetically favored \cite{ItagakiPRC2001,ItagakiPRL2004}. A number of studies have discussed the role of valence neutrons in carbon isotopes  \cite{ItagakiPRC2001,ItagakiPRC2006,MaruhnNPA2010,SuharaPRC2010,BabaPRC2014,EbranPRC2014,BabaPRC2016,ZhaoPRL2015,RenSCPMA2019,RenPLB2020,FreerPRC2014,LiPRC2017}.
In Ref.~\cite{ZhaoPRL2015}, both mechanisms have been considered simultaneously, and it has been shown that the valence neutron orbitals change from the $\pi$ orbital to the $\sigma$ orbital with increasing rotational frequency for $^{15-18}$C, which enhances the stability of linear chain configurations.  However, the rotational effect on the stability of linear chain configurations for proton-rich nuclei has not been discussed so far. Thus it will be interesting to investigate the effects of rotation and valence protons in proton-rich nuclei, eventual differences with respect to neutron-rich nuclei, as well as extend such studies to heavier nuclear system.
			
Various microscopic models have been developed to investigate linear chains of cluster structures. In addition to conventional approaches, such as the resonating group method (RGM) \cite{WheelerPR1937}, the generator coordinate method(GCM) \cite{DescouvemontNPA2002}, the molecular orbital (MO) model \cite{MichioPOTP1981,OkabePOTP1977,ItagakiPRC2000}, and the antisymmetrized molecular dynamics (AMD) method \cite{KanadaPRC1995,KanadaPOTP2001}, models based on nuclear density functional theory (DFT) provide a successful description of the linear chain structures \cite{Bender2003,Ichikawa2011,Yao2014,Iwata2015,MaruhnNPA2010,ZhaoPRL2015,RenSCPMA2019,RenPLB2020}. Nuclear DFT presents a self-consistent framework in which phenomena related to cluster structures can be investigated without {\em `a priori'} assuming the existence of $\alpha$-clusters. The relativistic extension of nuclear DFT, that is, the covariant DFT naturally includes the spin degree of freedom of the nucleon, the spin-orbital interaction~\cite{RenPRCR2020}, and nuclear currents, which is particularly important for the description of collective rotations~\cite{AfanasjevPRC2010,MengFront2013}.
			
In this work, the rotational effects in molecular linear $\alpha$-chain nuclei characterized by the alignment of valence nucleons, are investigated using the three-dimensional (3D) lattice cranking covariant density functional theory (CDFT), in which the variational collapse is avoided by the inverse Hamiltonian method~\cite{HaginoPRC2010} and the Fermion doubling problem is solved by spectral method~\cite{RenPRC2017}. The paper is organized as follows. In Sec.~\ref{Sec2}, we briefly review the basic formalism of the 3D lattice cranking CDFT, together with the localization function. Numerical details of the calculation and the principal results for rotating $\alpha$-chain nuclei with valence nucleons are presented and discussed in Sec.~\ref{Sec3}. Finally, a brief summary and outlook for future studies are included in Sec.~\ref{Sec4}.

\section{Theoretical framework}\label{Sec2}
\subsection{Cranking covariant density functional theory in 3D lattice space}

The framework of CDFT can be based on an effective nuclear Lagrangian density, that describes the strong and electromagnetic interactions between nucleons by the exchange of mesons and the photon, respectively~\cite{Meng2016}
\begin{align}\label{eq1}
\mathcal{L}&=\bar{\psi}\left[i\gamma^{\mu}\partial_{\mu}-M-g_{\sigma}\sigma-g_{\omega}\gamma^{\mu}
\omega_{\mu}-g_{\rho}\gamma^{\mu}\vec{\tau}\cdot\vec{\rho}_{\mu}\right.\nonumber\\
&\left.-e\gamma^{\mu}\frac{1-\tau_3}{2}A_{\mu}\right]\psi\nonumber\\
&+\frac{1}{2}\partial^{\mu}\sigma\partial_{\mu}\sigma-\frac{1}{2}m_{\sigma}^2\sigma^2-\frac{1}{4}\Omega^{\mu\nu}\Omega_{\mu\nu}+\frac{1}{2}m_{\omega}^2\omega^{\mu}\omega_{\mu}\nonumber\\
&-\frac{1}{4}\vec{R}^{\mu\nu}\cdot\vec{R}_{\mu\nu}+\frac{1}{2}m_{\rho}^2\vec{\rho}^{\mu}\cdot\vec{\rho}_{\mu}-\frac{1}{4}F^{\mu\nu}F_{\mu\nu},
\end{align}
where $M$, $m_{\sigma}$, $m_{\omega}$, and $m_{\rho}$ are the masses of the nucleon, $\sigma$ meson, $\omega$ meson, and $\rho$ meson, respectively. $g_{\sigma}$, $g_{\omega}$, and $g_{\rho}$ are the corresponding couplings for the mesons to the nucleon and, in general, these are functions of the nucleon density. $\Omega^{\mu\nu}$, $\vec{R}^{\mu\nu}$, and $F^{\mu\nu}$ are the field tensors of the vector fields $\omega$, $\rho$, and the photon.

To describe nuclear rotations in the cranking approximation, the effective Lagrangian density of Eq.~(\ref{eq1}) is transformed into a rotating frame with a constant rotational frequency around a given rotational axis. Taking the $y$-axis as the axis of rotation, the single-nucleon equation of motion is derived from the Lagrangian in the rotating frame
\begin{align}\label{eq2}
\hat{h}'\psi_k=(\hat{h}_0-\omega\hat{j}_y)\psi_k=\epsilon'_k\psi_k,
\end{align}
where $\hat{h}_0$ is single-nucleon Hamiltonian,
\begin{align}
\hat{h}_0=\bm{\alpha}\cdot(\bm{p}-\bm{V})+\beta(m+S)+V_0,
\end{align}
Here $\hat{j_y}=\hat{l_y}+\frac{1}{2}\hat{\Sigma_y}$ is the y-component of the total angular momentum of the nucleon spinor, and $\epsilon'_k$ is the single-particle Routhian. The relativistic scalar meson field $S$ and vector meson field $V_{\mu}$ are related
in a self-consistent way to the nucleon densities and current distributions. By solving the cranking Dirac equation~(\ref{eq2}) self-consistently, one obtain the single-particle Routhians, the expectation values of the angular momentum, quadrupole moments, etc.

In this work, the cranking Dirac equation is solved in 3D lattice space. The main challenges one encounters are the variational collapse and the fermion doubling problem, which are solved by the inverse Hamiltonian method~\cite{HaginoPRC2010} and spectral method~\cite{RenPRC2017}, respectively.  For details, we refer the reader to Ref.~\cite{RenPRC2017}. A damping function is introduced in the cranking term to remove the unphysical continuum effect on single-particle Routhians with large angular momenta~\cite{RenSCPMA2019}. The cranking term $-\omega\hat{j}_y$ in Eq.~(\ref{eq2}) is replaced by
\begin{align}
f_D(r)(-\omega\hat{j}_y)f_D(r),
\end{align}
where the damping function is of a Fermi-type determined by two parameter $r_D$ and $a_D$:
\begin{align}
f_D(r)=\frac{1}{1+e^{r-r_D}/a_D}.
\end{align}

\subsection{Localization function}

The conditional probability of finding a nucleon within a distance $\delta$ from a given nucleon at point $\bm{r}$ with the same spin $\sigma$ ($=\uparrow$ or $\downarrow$)  and isospin $q$ ($=n$ or $p$) quantum numbers is,
\begin{equation}
R_{q\sigma}(\bm{r},\delta)\approx{1\over 3}\left(\tau_{q\sigma}-{1\over 4}\frac{|\bm{\nabla}\rho_{q\sigma}|^2}{\rho_{q\sigma}}-\frac{\bm{j}^2_{q\sigma}}{\rho_{q\sigma}}\right)\delta^2+\cal{O}(\delta^3),
\end{equation}
where $\rho_{q\sigma}$, $\tau_{q\sigma}$, $\bm{j}_{q\sigma}$, and $\bm{\nabla}\rho_{q\sigma}$ denote the particle density, kinetic energy density, current density, and density gradient, respectively, and are completely determined by the self-consistent mean-field single-particle states.
From the conditional probability, the nucleon localization function can be derived as~\cite{ReinhardPRC2011,EbranJPG2017,RenNPA2020},
\begin{equation} \label{nlf}
C_{q\sigma}(\bm{r})=\left[1+\left(\frac{\tau_{q\sigma}\rho_{q\sigma}-{1\over 4}|\bm{\nabla}\rho_{q\sigma}|^2-\bm{j}^2_{q\sigma}}{\rho_{q\sigma}\tau^\mathrm{TF}_{q\sigma}}\right)^2\right]^{-1} ,
\end{equation}
where $\tau^\mathrm{TF}_{q\sigma}={3\over 5}(6\pi^2)^{2/3}\rho_{q\sigma}^{5/3}$ is the Thomas-Fermi kinetic energy density. The function $C_{q\sigma}(\bm{r})$ is normalized and provides a dimensionless measure of nucleon localization.
For homogeneous nuclear matter $\tau = \tau^\mathrm{TF}_{q\sigma}$, the second and third term in the numerator vanish,
and $C_{q\sigma} = 1/2$. In the other limit $C_{q\sigma} (\bm{r}) \approx 1$
indicates that the probability of finding two nucleons with the same spin and isospin at the same point $\bm{r}$ is very small.
This is the case for the $\alpha$-cluster of four particles: $p \uparrow$,  $p \downarrow$, $n \uparrow$,
and $n \downarrow$, for which all four nucleon localization functions $C_{q\sigma} \approx 1$.

To emphasize localization inside a nucleus and avoid numerical instabilities in the outside region where densities are very small, a masking function is used to suppress the localization function in the region where this quantity is no longer relevant:
\begin{align}
C_{q\sigma}(\bm{r})\rightarrow C_{q\sigma}(\bm{r})\rho_{q\sigma}(\bm{r})/\max[\rho_{q\sigma}(\bm{r})].
\end{align}

For 3D lattice CDFT calculations that include rotations, spin is not a conserved quantity because of broken time-reversal symmetry. Since we are interested in the total localization of neutrons and protons, the average localization function is considered: $C_{q}^{\text{av}}=(C_{q\uparrow}+C_{q\downarrow})/2$.

\section{Results and discussion}\label{Sec3}

In this work, the density functional DD-ME2~\cite{LalazissisPRC2005} is employed in self-consistent cranking mean-field calculations. The number of grid points is 32 in the $x$ and $y$ directions, and 40 in the $z$ direction. The step size along all three axes is 0.8 fm. The parameters $r_D=11$ fm and $a_D=0.2$ fm of the Fermi damping function ensure that the convergence is achieved. The parameters used in the inverse Hamiltonian method are same as in Ref.~\cite{RenSCPMA2019}

The present study of cluster structures starts with $^{12}$C. In the first step, Eq.~(\ref{eq2}) is solved iteratively
at zero rotational frequency, by assuming the initial potential with a very large prolate quadrupole deformation. A shape constrained self-consistent solution with a 3$\alpha$ linear chain configuration is thus obtained. Using this potential as the initial potential, cranking CDFT calculations in 3D lattice space are  performed self-consistently for $^{12}$C, $^{16}$C, and $^{16}$Ne at different rotational frequencies. The linear chain configurations at low rotational frequencies are obtained in such a way that the proton configuration for $^{16}$C and the neutron configuration for $^{16}$Ne, are constrained to reproduce the linear 3$\alpha$-chain configuration of $^{12}$C.

\begin{figure}[t]
	\centering
	\includegraphics[width=0.5\linewidth]{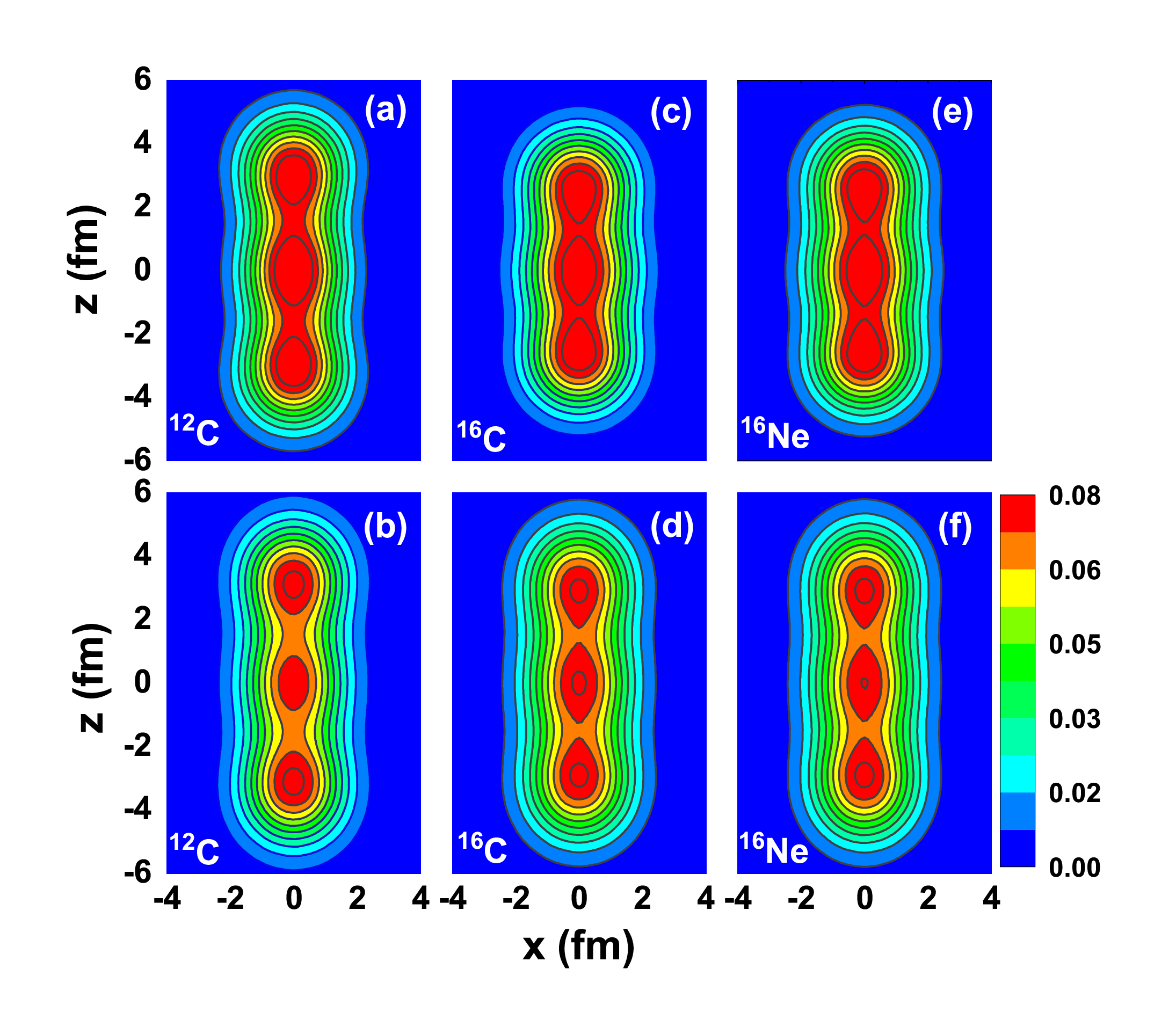}
	\caption{Proton density distributions  in the $x$-$z$ plane for $^{12}$C, $^{16}$C and neutron density distributions in the $x$-$z$ plane for $^{16}$Ne at the rotational frequencies $\hbar\omega=0.0$ MeV (a),(c),(e) and $\hbar\omega=3.0$ MeV (b),(d),(f).}
	\label{fig1}
\end{figure}

Figure~\ref{fig1} displays the proton density distributions for $^{12}$C and $^{16}$C, as well as the  neutron density distributions for $^{16}$Ne, at rotational frequencies $\hbar\omega=0.0$ MeV and $\hbar\omega=3.0$ MeV. The 3$\alpha$-chain is clearly seen in all cases, and it is interesting to note how these structures are elongated along the $z$ axis by centrifugal stretching, when increasing the rotational frequency to $\hbar\omega=3.0$ MeV.

\begin{figure*}[t]
	\centering
	\includegraphics[width=1\linewidth]{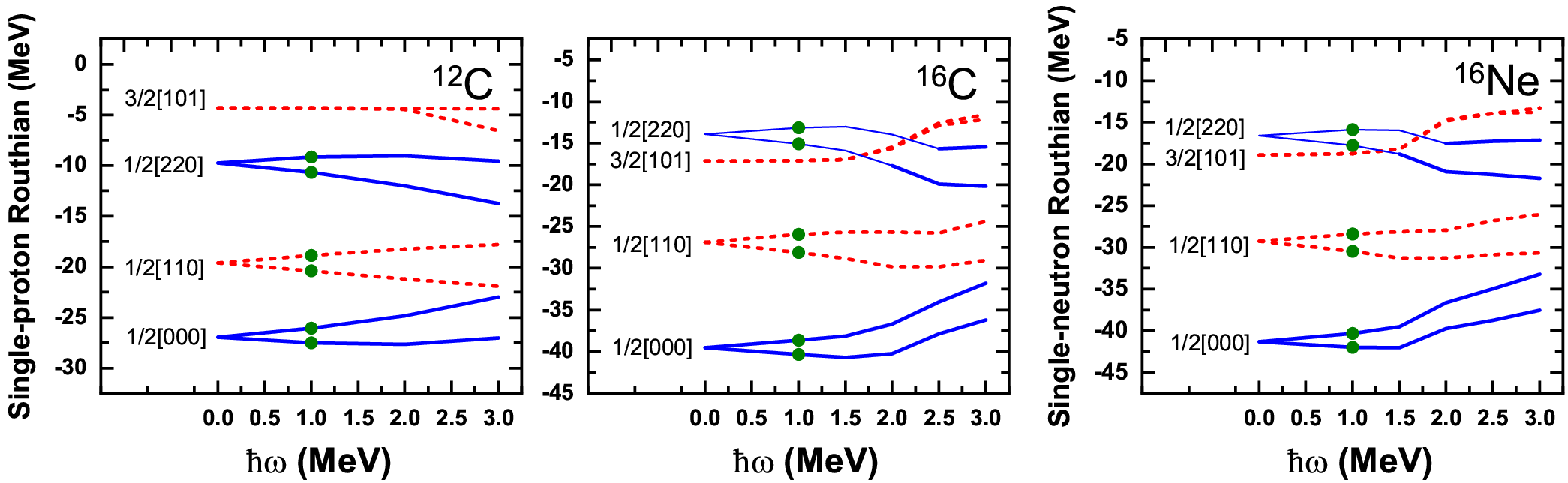}
	\caption{ Single-proton Routhians as functions of the rotational frequency for  $^{12}$C and   $^{16}$C, and single-neutron Routhians for  $^{16}$Ne. Each orbital is labeled by the corresponding Nilsson quantum number of its maximal component. The solid and dashed lines correspond to single-particle states with positive and negative parity, respectively. Circles denote the occupied orbitals.
	}
	\label{fig2}
\end{figure*}

Although the 3$\alpha$ linear chain structure persists in the mirror nuclei $^{16}$C and $^{16}$Ne, these configurations are not necessarily stable at all rotational frequencies. In Fig.~\ref{fig2} we plot the single-particle Routhians as functions of the rotational frequency for $^{12}$C,  $^{16}$C, and $^{16}$Ne. Each level is labeled by the corresponding Nilsson quantum number of its maximal component, and levels of positive and negative parity are denoted by solid and dashed lines, respectively. All levels are doubly degenerate at $\hbar\omega=0.0$ MeV because of time-reversal symmetry, and split into two levels each as the rotational frequency increases.

For the nucleus $^{12}$C, the occupied proton states are always the lowest-energy levels and this configuration does not change with increasing rotational frequency.
A different situation occurs for $^{16}$C, with four more valence neutrons. As noted above, to obtain the linear 3$\alpha$-chain configuration at low rotational frequency, the occupation of proton levels is constrained in such a way that the level 3/2[101] is unoccupied, even though it is lower in energy than the occupied levels originating from the Nilsson level 1/2[220]. This means that the last two protons in the level 1/2[220] can occupy the level 3/2[101] to lower the energy and, therefore, the 3$\alpha$ linear chain proton configuration is not stable at small rotational frequencies. Note that the linear chain structure requires that the proton configuration for $^{16}$C does not change, thus the levels originating from 1/2[220] are denoted by the thin lines. With the rotational frequency increasing, the occupied levels 1/2[220] become lower in energy with respect to the level 3/2[101]. The first level crossing occurs at $\hbar\omega \approx 1.7$ MeV, and the 3$\alpha$ linear chain proton configuration eventually stabilizes at rotational frequency $\hbar\omega \approx 2.2$  MeV.

The nucleus $^{16}$Ne has four more valence protons compared to $^{12}$C and, analogous to the proton levels in $^{16}$C, the corresponding first neutron level crossing occurs already at $\hbar\omega \approx 1.3$ MeV. One notices, however, that
the 3$\alpha$ linear chain neutron configuration is stabilized at rotational frequencies $\hbar\omega \approx 1.7$ MeV. When compared to $^{16}$C, it appears that the 3$\alpha$ linear chain configuration could be stabilized at lower rotational frequency.

\begin{figure}[t]
	\centering
	\includegraphics[width=0.5\linewidth]{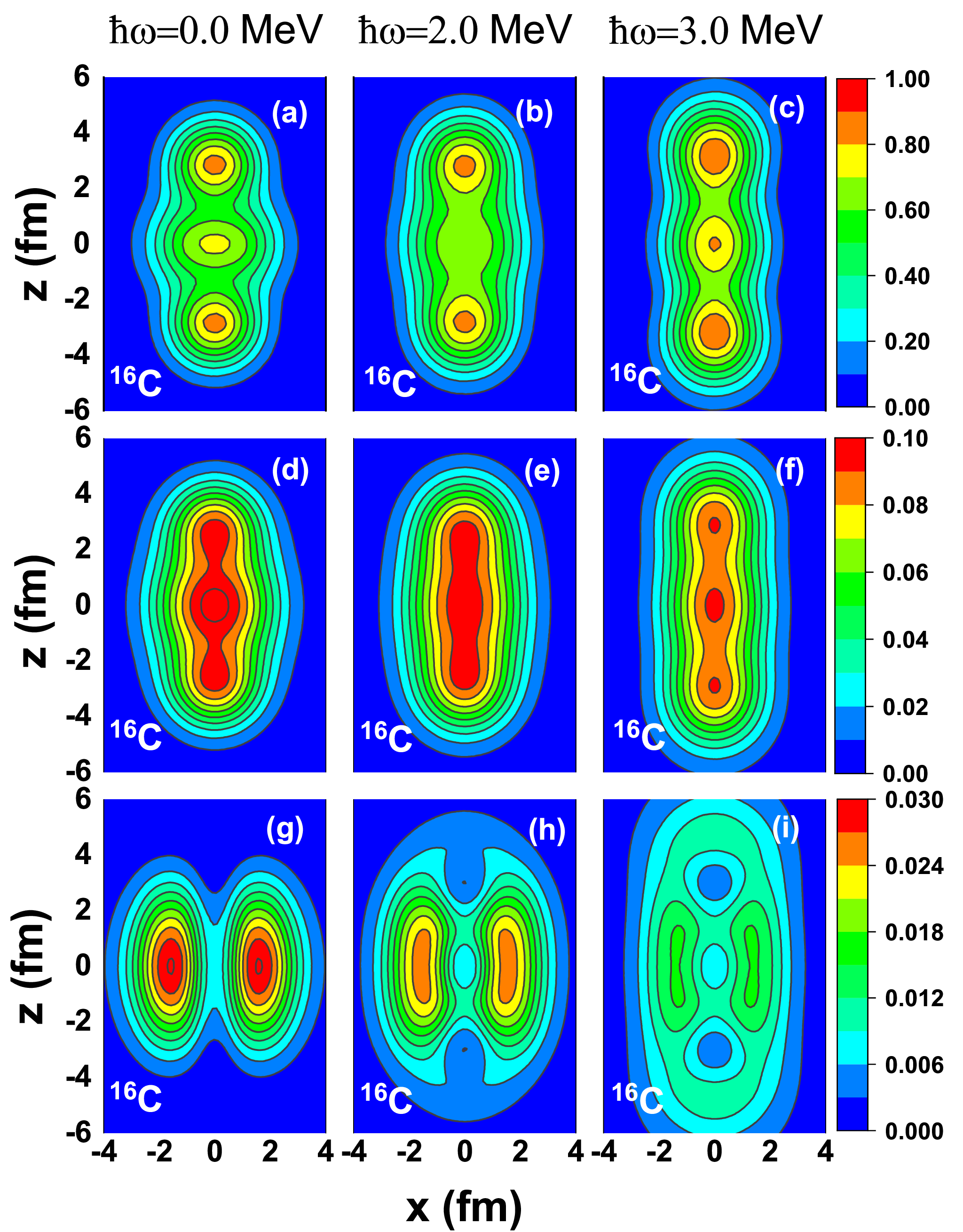}
	\caption{Neutron localization function (a),(b),(c), neutron density distributions (d),(e),(f), valence particle density distributions (g),(h),(i) in the $x$-$z$ plane for  $^{16}$C at the rotational frequencies $\hbar\omega$=0.0, 2.0, and 3.0 MeV, respectively.}
	\label{fig3}
\end{figure}

The stability of the linear 3$\alpha$-chain configuration in the mirror nuclei $^{16}$C and $^{16}$Ne is affected by the valence particles. In Fig.~\ref{fig3} we plot the average neutron localization function ($C_{n}^{\text{av}}$), the neutron density distributions, and the valence neutron density distributions for  $^{16}$C at  rotational frequencies: $\hbar\omega$=0.0, 2.0, and 3.0 MeV, respectively.
At $\hbar\omega=0.0$ MeV, $C_{n}^{\text{av}}$ displays two symmetric regions of pronounced  localization, but the neutron density appears to be less localized in the central region. This is because the valence neutrons exhibit an oblate distribution with respect to the $z$ axis (panel (g)), indicating that  the valence neutrons occupy the $\pi$ orbital. The occupation of the $\pi$ orbital contributes to the neutron density in the $z=0$ plane, thus reducing the localization of the central $\alpha$ particle.
With increasing rotational frequency, the valence neutrons density changes from an oblate distribution to a more prolate distribution. At $\hbar\omega = 2.0$ MeV, the valence neutrons partly occupy the $\sigma$ orbital. However, there is no clear localization of the 3$\alpha$ linear chain structure in the map of $C_{n}^{\text{av}}$ or the neutron density distributions. As noted above, the 3$\alpha$ linear chain proton configuration is not yet stabilized, according to the single-proton Routhians for $^{16}$C in Fig.~\ref{fig2}.

At  $\hbar\omega=3.0$ MeV, the valence neutrons display a prolate distribution with respect to the $z$ axis. This means that the valence neutrons change the occupation from the $\pi$ orbital to the $\sigma$ orbital, thus strengthening the stability of the 3$\alpha$ linear chain structure. The 3$\alpha$ linear chain structure is clearly seen in the map of $C_{n}^{\text{av}}$ and the neutron density distribution. Furthermore, the proton configuration that determines the 3$\alpha$ structure has been stabilized at this rotational frequency (cf. the single-proton Routhians in Fig.~\ref{fig2}). The localization of the $\alpha$-cluster in the center is slightly different from the peripheral ones because of the contribution of the valence neutrons.

\begin{figure}[t]
	\centering
	\includegraphics[width=0.5\linewidth]{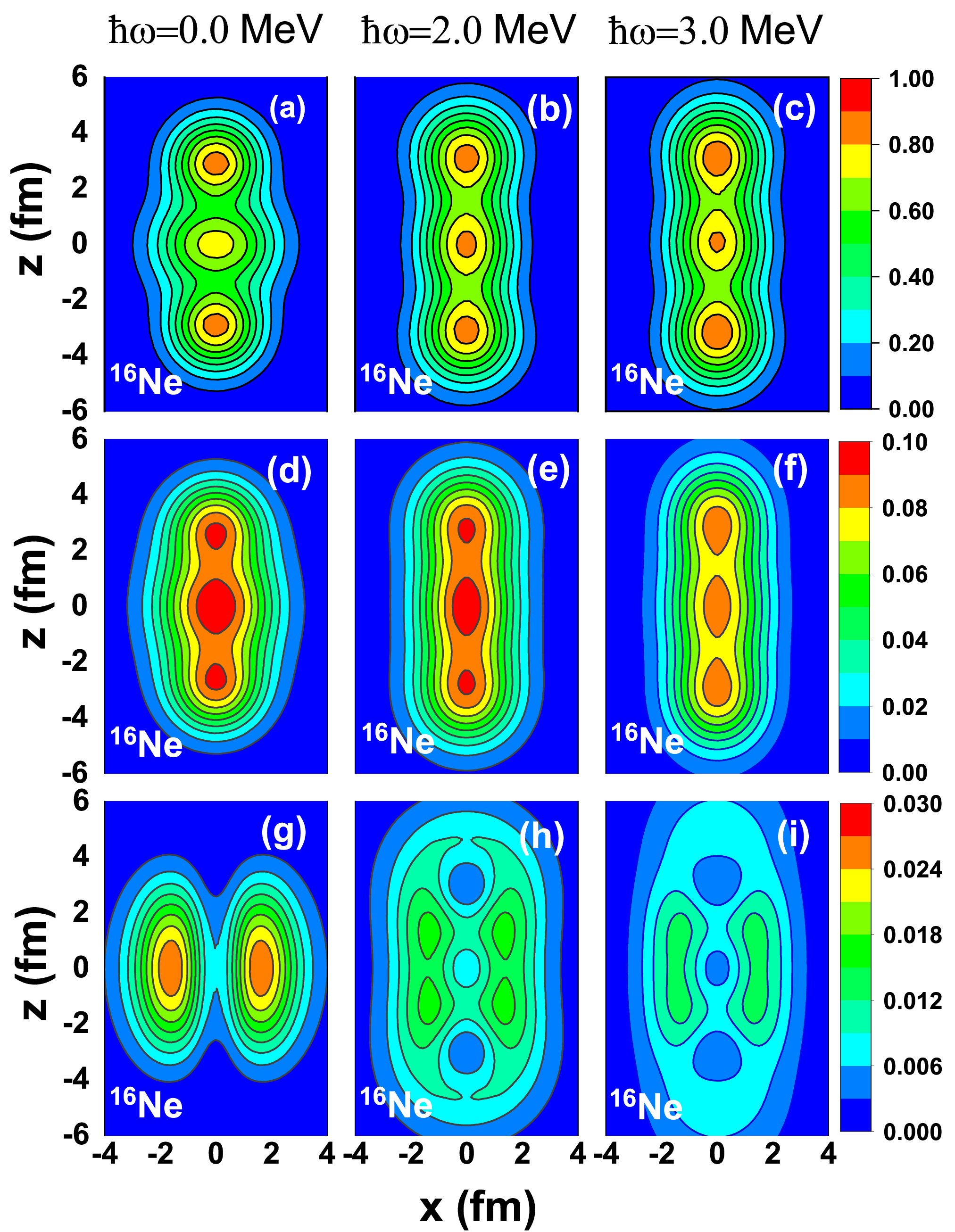}
	\caption{Proton localization function (a),(b),(c), proton density distributions (d),(e),(f), valence particle density distributions (g),(h),(i) in the $x$-$z$ plane for  $^{16}$Ne at the rotational frequencies $\hbar\omega$=0.0, 2.0, and 3.0 MeV, respectively.}
	\label{fig4}
\end{figure}

Figure~\ref{fig4} displays the corresponding proton average localization function ($C_{p}^{\text{av}}$), proton density distributions, and the valence proton density distributions for $^{16}$Ne. Similar to $^{16}$C, the valence protons occupy the $\pi$ orbital and delocalize the 3$\alpha$ linear chain structure at $\hbar\omega=0.0$ MeV. When the rotational frequency is increased to $\hbar\omega=2.0$ MeV, the valence proton density already exhibits a prolate distribution along the $z$ axis. The linear 3$\alpha$-chain structure is clearly seen in the map of $C_{p}^{\text{av}}$ and the proton density distribution. It appears that the valence protons in $^{16}$Ne transition from the $\pi$ orbital to the $\sigma$ orbital at lower rotational frequency compared to the valence neutrons in $^{16}$C. Consequently, the linear 3$\alpha$-chain configuration gets stabilized at a lower rotational frequency in the 3$\alpha + 4 p$ system.

\begin{figure}[t]
	\centering
	\includegraphics[width=0.8\linewidth]{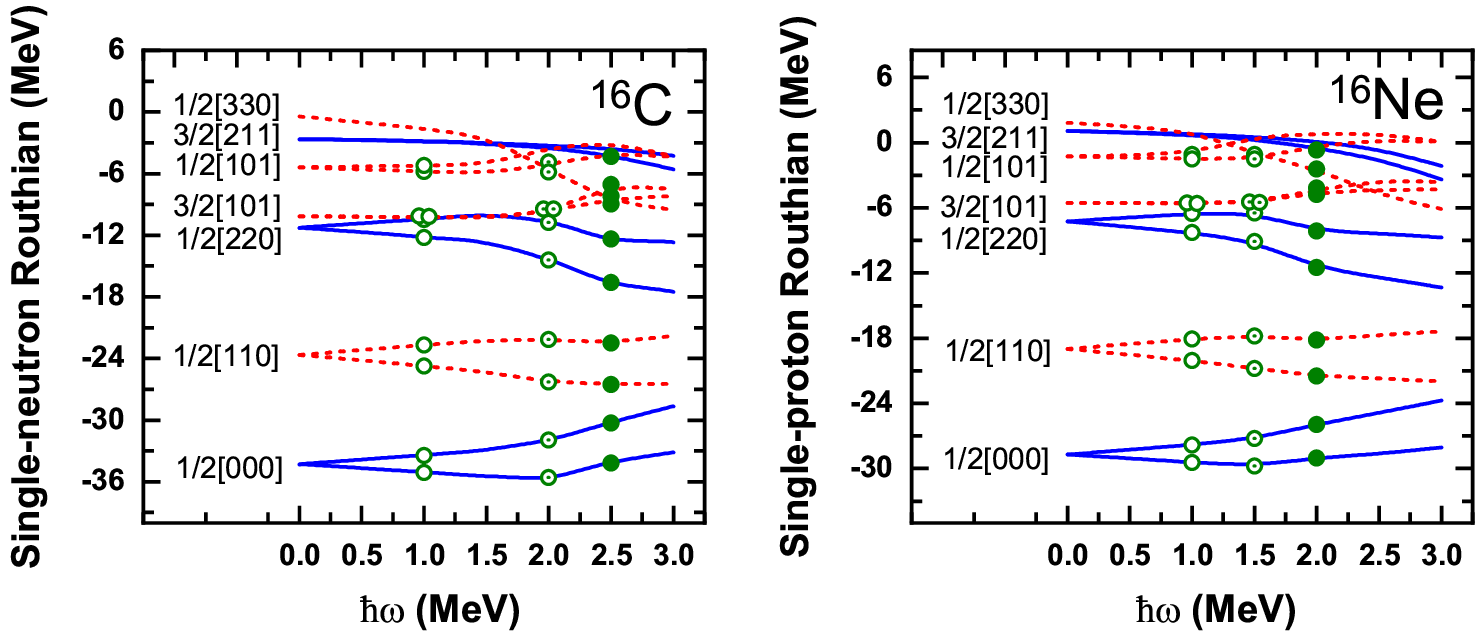}
	\caption{Single-neutron Routhians for $^{16}$C, and single-proton Routhians for $^{16}$Ne, as functions of the rotational frequency. The open, open-dot, and filled circles denote the occupied orbitals before the level crossing, after the first crossing, and after the second crossing, respectively.}
	\label{fig5}
\end{figure}

The single-neutron Routhians for $^{16}$C, and single-proton Routhians for $^{16}$Ne are shown in Fig.~\ref{fig5}. One notices how the level originating from 1/2[330] decreases in energy as a function of rotational frequency and crosses the levels that correspond to 3/2[211] and 1/2[101]. The downsloping level becomes occupied at higher frequencies. For $^{16}$C, in particular, we see that at $\hbar\omega \approx 1.7$ MeV a valence neutron firstly changes the occupation from the level originating in 1/2[101] to the level corresponding to 1/2[330], and afterwards another valence neutron occupies a level corresponding to 3/2[211] at $\hbar\omega \approx 2.5$ MeV. A similar behavior is also observed for $^{16}$Ne, but the valence protons change the occupation from $\pi$  to $\sigma$ molecular orbitals at a lower rotational frequency.

\begin{figure}[t]
  \centering
\includegraphics[width=0.5\linewidth]{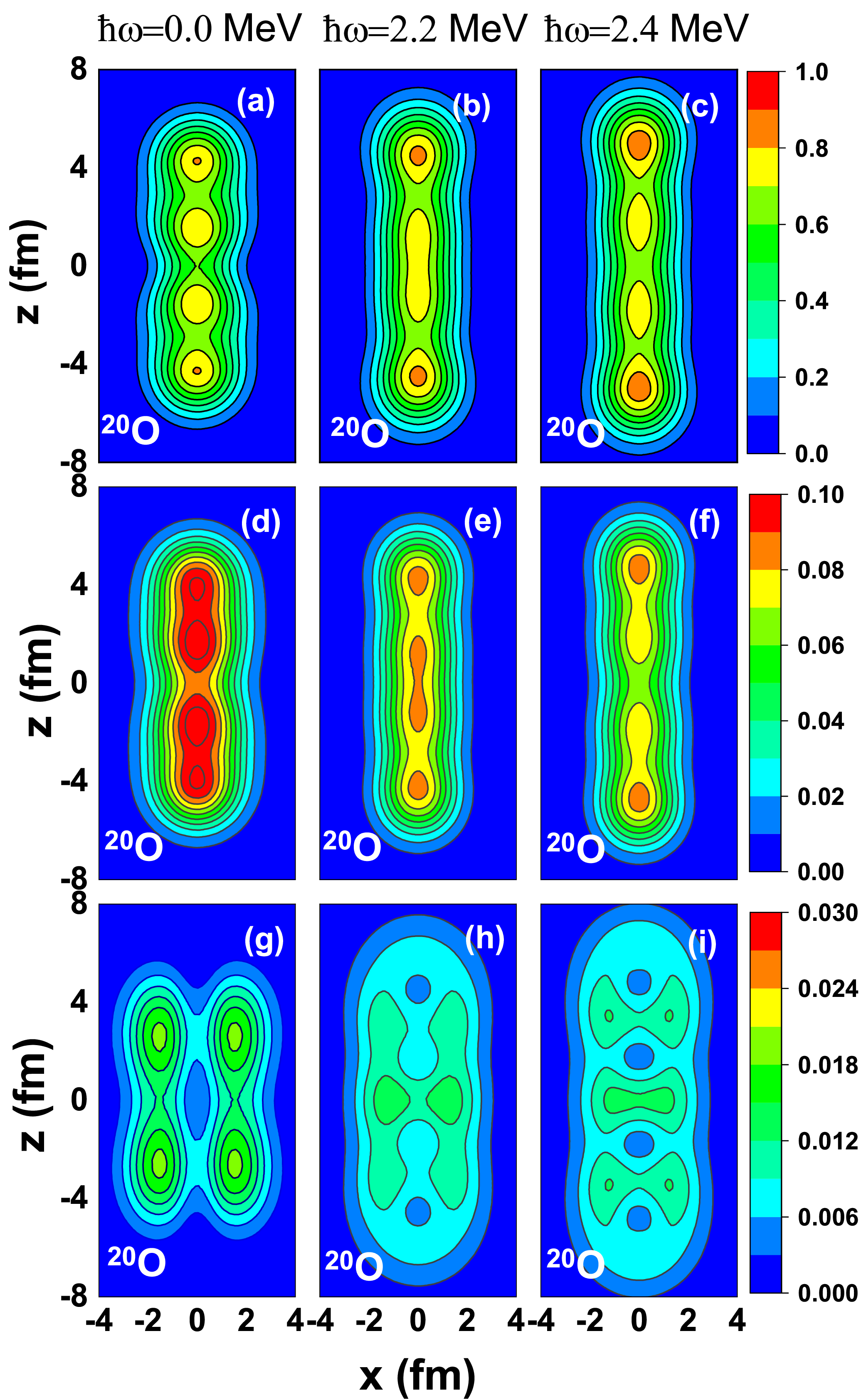}
\caption{Same as in the caption to Fig.~\ref{fig3} but for  $^{20}$O at the rotational frequencies $\hbar\omega$=0.0, 2.2, and 2.4 MeV.}
  \label{fig6}
\end{figure}

\begin{figure}[t]
	\centering
	\includegraphics[width=0.5\linewidth]{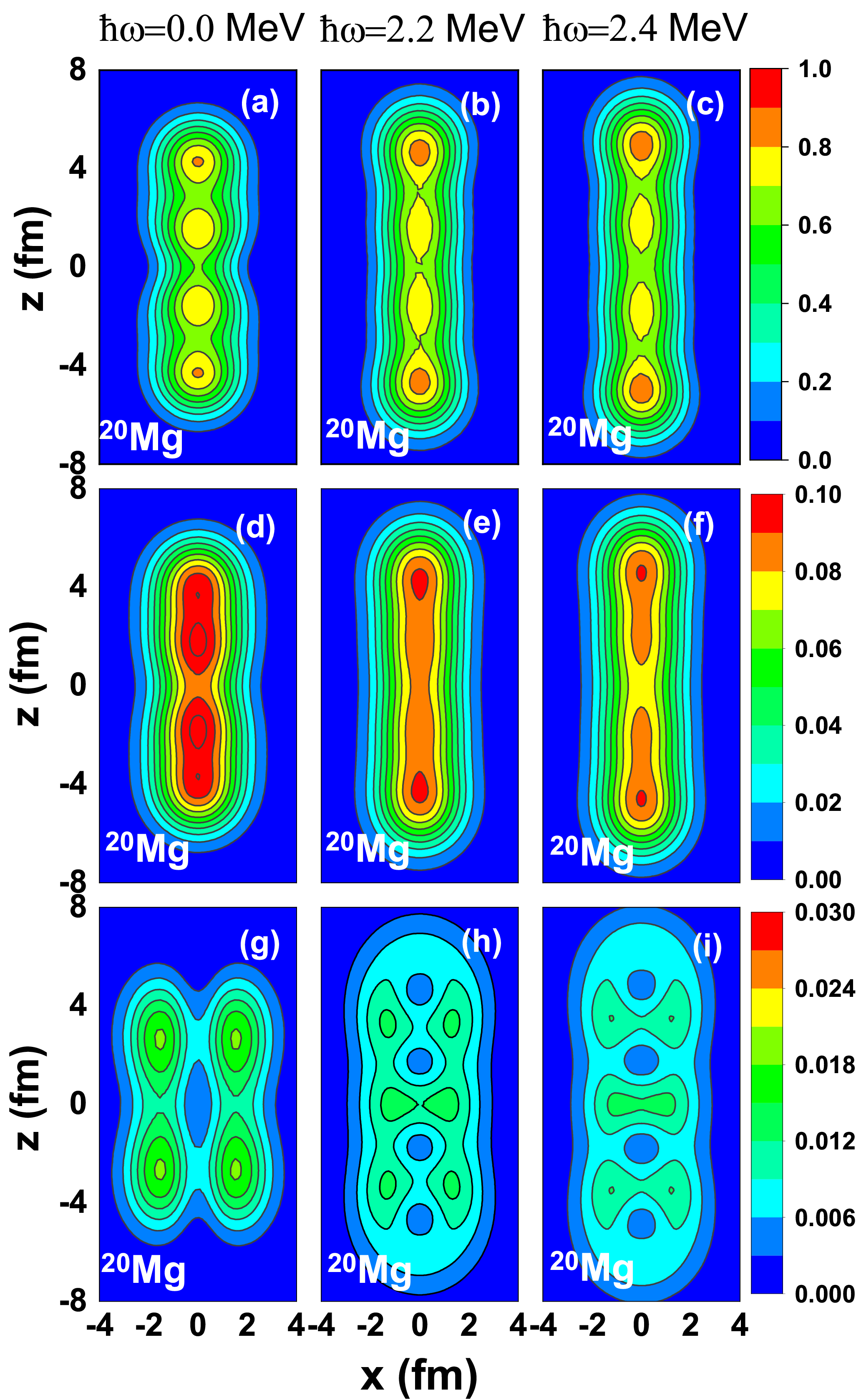}
	\caption{Same as in the caption to Fig.~\ref{fig4} but for $^{20}$Mg at the rotational frequencies $\hbar\omega$=0.0, 2.2, and 2.4 MeV.}
	\label{fig7}
\end{figure}

One expects that a similar effect might occur in heavier systems. Therefore, we have carried out the corresponding  calculations for the mirror nuclei $^{20}$O and $^{20}$Mg. The results are shown in Figs.~\ref{fig6} and \ref{fig7}. Both for the mirror nuclei $^{20}$O and $^{20}$Mg, the valence particle densities in the lower panels exhibit distributions that are characteristic for the occupation of the $\pi$ orbital at $\hbar\omega=0.0$ MeV. At $\hbar\omega \approx 2.2$ MeV, for the nucleus $^{20}$O, valence neutrons start to change the occupation from the $\pi$ orbital to the $\sigma$ orbital. However, the contribution of the $\sigma$ orbital is still not dominant. As a result, the neutron density extends along the $z$ direction, and the two $\alpha$-clusters in the central region are difficult to identify from the map of $C_n^{\text{av}}$ and the neutron density distribution. For the nucleus $^{20}$Mg, already at this frequency the valence protons predominantly occupy the $\sigma$ orbital, and the 4$\alpha$ linear chain structure is significantly more localized in the map of $C_p^{\text{av}}$. This shows that the valence protons change the occupation from the $\pi$ orbital to the $\sigma$ orbital at lower rotational frequency compared to the valence neutrons that supplement the 4$\alpha$ system in $^{20}$O.
At $\hbar\omega=2.4$ MeV, four nodes along the $z$ axis are clearly identified in the density distributions of the valence particles both for $^{20}$O and $^{20}$Mg, and the 4$\alpha$ linear chain structure is more pronounced in the maps of the localization function.

We have also analyzed the single-proton and the single-neutron Routhians as functions of rotational frequency for the mirror nuclei $^{20}$O and $^{20}$Mg, respectively. For $^{20}$O, we find that the 4$\alpha$ linear chain neutron configuration stabilizes at $\hbar\omega \approx 2.4$ MeV, while for $^{20}$Mg the 4$\alpha$ linear chain proton configuration appears stable already at $\hbar\omega \approx 2.2$ MeV. Therefore, with increasing  rotational frequency, the linear chain configuration is stabilized at lower rotational frequency for the 4$\alpha+4 p$ system. The valence protons in $^{20}$Mg change the occupation from the $\pi$ orbital to the $\sigma$ orbital at lower frequency compared to the valence neutrons in $^{20}$O.

\begin{figure}[t]
	\centering
	\includegraphics[width=0.7\linewidth]{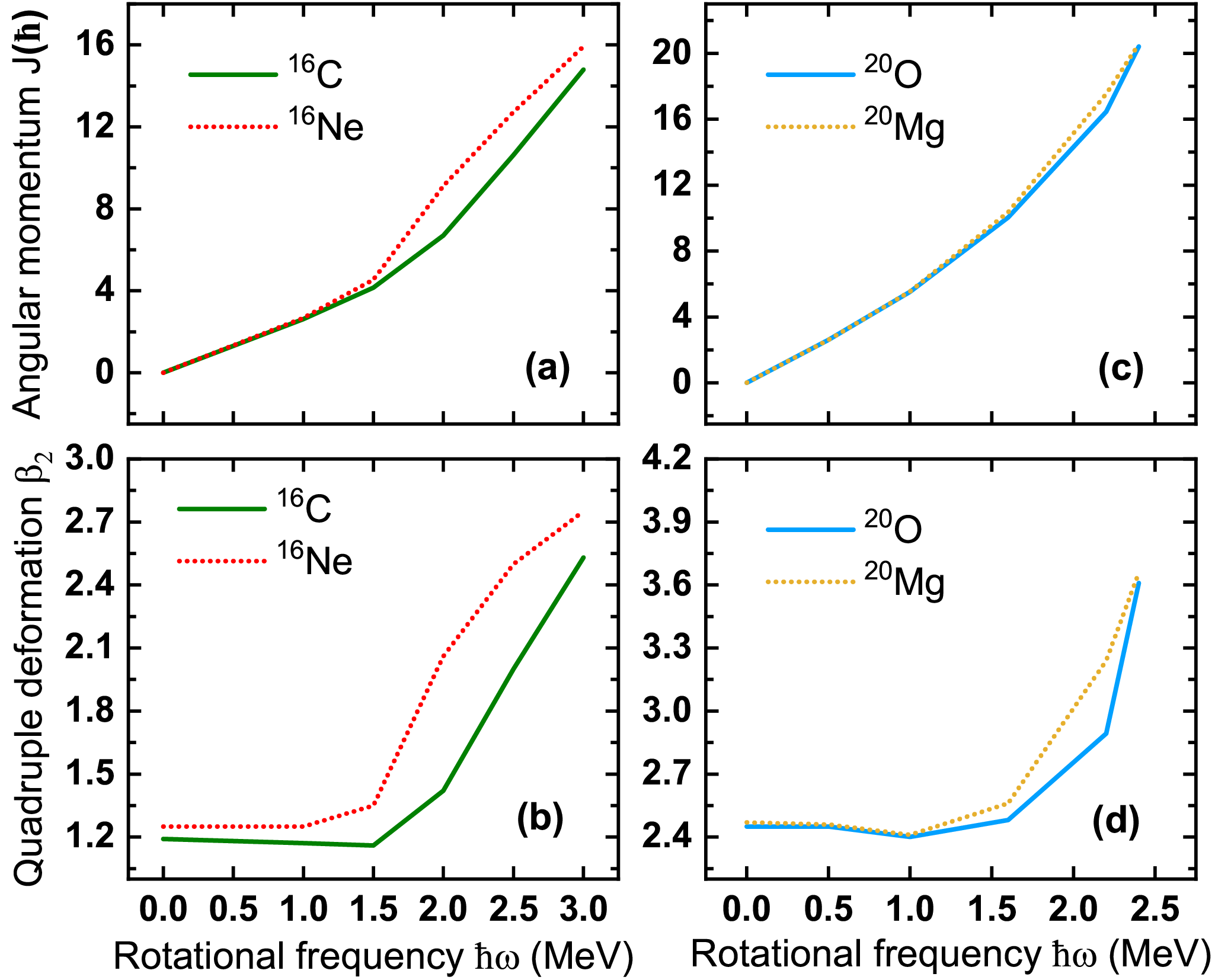}
	\caption{Angular momenta (a),(c) and quadrupole deformation $\beta_2$ (b),(d) as functions of rotational frequency for $^{16}$C, $^{16}$Ne, $^{20}$O, and $^{20}$Mg.}
	\label{fig8}
\end{figure}

The phenomenon described above is also illustrated in the alignment of the angular momentum and the evolution of quadrupole deformation with rotational frequency. Figure~\ref{fig8} displays the angular momentum and the quadrupole deformation $\beta_2$ for  $^{16}$C, $^{16}$Ne, $^{20}$O, and $^{20}$Mg, as functions of rotational frequency. Initially, in all cases the angular momentum increases linearly with rotational frequency, which means that the moments of inertia are nearly constant. This is because the single-nucleon configurations do not change at low rotational frequencies.

The slope suddenly changes at $\hbar\omega \approx 1.5$ MeV for $^{16}$Ne and $\hbar\omega \approx 2.0$ MeV for $^{16}$C (panel (a)), and this indicates a structural change with increasing angular momentum. At these rotational frequencies the valence nucleons begin to change their occupation from the $\pi$ orbital to the $\sigma$ orbital.  Note that the alignment occurs at lower rotational frequency in $^{16}$Ne. The effect is also clearly seen, and even more pronounced, in the plot of the quadrupole deformation parameter $\beta_2$ as function of the rotational frequency (panel (b)). The quadrupole moment of $^{16}$C is almost constant up to $\hbar\omega \approx 1.5$ MeV, where it exhibits an abrupt increase. The corresponding sharp rise of the quadrupole deformation in $^{16}$Ne occurs already at $\hbar\omega \approx 1.5$ MeV. A similar behavior, but perhaps with a less pronounced difference between systems with valence neutrons and protons, is also seen for the mirror nuclei
 $^{20}$Mg and $^{20}$O (panels (c) and (d)).

\begin{figure}[t]
	\centering
	\includegraphics[width=0.5\linewidth]{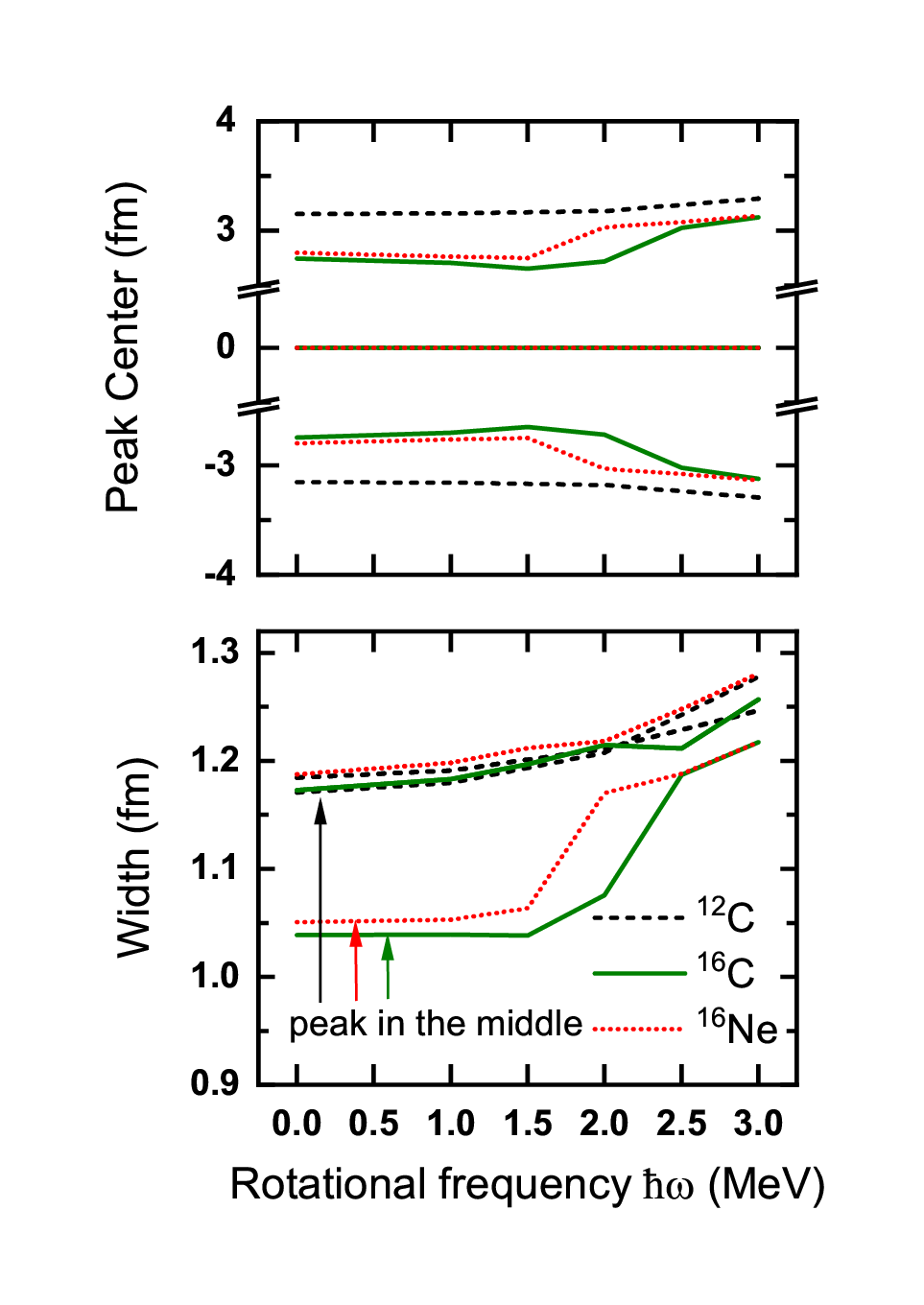}
	\caption{Location of the peak (upper panel), and the width (lower panel) of each $\alpha$-like cluster in $^{12}$C, $^{16}$C, and $^{16}$Ne, obtained by fitting the $3\alpha$ core density along $z$ direction by a linear combination of Gaussian functions, as functions of rotational frequency.}
	\label{fig9}
\end{figure}

An interesting point, that has not been considered in previous studies of molecular bonding in nuclei with cluster structures, is the effect of valence nucleons on the characteristics of the clusters (relative positions, cluster size).
In all cases considered in the present study, one notices that the central cluster structures differ from the outer ones which, generally, exhibit more localization. This difference, however, depends on the rotational frequency. Here we analyze this behavior for the $\alpha$ linear chain configurations in $^{12}$C, $^{16}$C and $^{16}$Ne. The $3\alpha$-like density distributions of the core nucleons, that is, excluding the valence particles, are fitted along $z$ direction (the density is integrated in the $x-y$ plane) by a linear combination of Gaussian functions:
\begin{align}
\rho=\sum_{i=1}^{3}\rho_i \exp\left[-\frac{(z-z_i)^2}{2w_i^2}\right]
\end{align}
where $\rho_i, z_i, w_i$ ($i=1,2,3$) denote the amplitude, the peak location, and the width, respectively of each $\alpha$-like cluster. The results are shown in Fig.~\ref{fig9}. In the upper panel we plot the
$z$-coordinates of the peak of each cluster. On the one hand, for $^{12}$C the position of the peripheral peaks hardly changes with rotational frequency, and only for $\hbar\omega > 2$ MeV one notices the effect of centrifugal stretching. For the mirror nuclei $^{16}$C and $^{16}$Ne, on the other hand, the positions of the outer peaks are much closer to the central one at low rotational frequencies. This is the effect of the valence nucleons occupying the $\pi$ orbital, and thus providing additional attraction for the outer $\alpha$ clusters. The transition of the valence nucleons from the $\pi$ to the $\sigma$ orbital at
$\hbar\omega \approx 1.5$ MeV for $^{16}$Ne and $\hbar\omega \approx 2.0$ MeV for $^{16}$C, weakens the bonding of the outer $\alpha$ clusters and the location of their peaks approaches that of the corresponding $\alpha$ particles in $^{12}$C. Moreover, the repulsive effect of the Coulomb interaction between protons stretches the linear chain configuration for proton-rich nuclei, and the distance between cluster peaks for $^{16}$Ne is always longer than for $^{16}$C. This mechanism makes it easier for proton-rich nuclei to stabilize the linear chain configuration.

The widths of the fitted Gaussian functions are shown, as functions of rotational frequency, in the lower panel of Fig.~\ref{fig9}. As one would expect, there is virtually no difference between the widths of the  central and outer $\alpha$ clusters in $^{12}$C for $\hbar\omega \leq 2$ MeV, and only a small effect of centrifugal stretching in the interval $2 \leq \hbar\omega \leq 3$ MeV. In contrast, for $^{16}$C and $^{16}$Ne one notices a pronounced reduction of the Gaussian width that corresponds to the central
$\alpha$ cluster. It appears as if the $\alpha$ cluster in the middle is squeezed by the outer ones that are more strongly bound by the valence particles occupying the $\pi$ orbital. Therefore, the transition from the $\pi$ to the $\sigma$ orbital not only releases the outer $\alpha$ clusters, but also results in the increase of the width of the central cluster peak which, after the transition, approaches values that are characteristic for the peripheral clusters, and the three $\alpha$ clusters in $^{12}$C.

\section{Summary}\label{Sec4}
By employing the 3D lattice cranking CDFT, we have investigated rotational effects in molecular linear $\alpha$-chain nuclei characterized by the alignment of valence nucleons. Starting from the 3$\alpha$ linear chain configuration in $^{12}$C, the structure of the mirror nuclei with four more neutrons ($^{16}$C) and four more protons ($^{16}$Ne) has been analyzed as a function of rotational frequency. At low frequency, or low angular momenta, the valence nucleons occupy the $\pi$ molecular orbital (perpendicular to the $z$ axis of the  3$\alpha$ linear chain core) and, although this configuration provides additional binding for the 3$\alpha$ structure, it cannot stabilize the 3$\alpha$ linear chain. With increasing rotational frequency, however, the valence nucleons transition from the $\pi$ orbital to the $\sigma$ molecular orbital
(parallel to the $z$ axis of the  3$\alpha$ linear chain core), and stabilize the 3$\alpha$ linear chain. At the same rotational frequency one finds a crossing between occupied and unoccupied Routhians of core nucleons with opposite isospin projection. The rather abrupt transitions is also reflected in the sudden change of alignment of the angular momentum and the quadrupole deformation of the nucleus.

An interesting result is that the valence protons in $^{16}$Ne change occupation from the $\pi$ to the $\sigma$ molecular orbital at lower rotational frequency compared to the valence neutrons in $^{16}$C. We have also verified that the same effect, although less pronounced, is found for the alignment of valence protons in $^{20}$Mg when compared to the four valence neutrons in $^{20}$O. The repulsive effect of the Coulomb interaction increases the distance between clusters along the z-axis for proton-rich nuclei, and thus stabilizes the linear chain configurations at a lower rotational frequency. Finally, we have investigated the effect of the alignment of valence nucleons on the relative positions and size of the three $\alpha$-clusters in the mirror nuclei $^{16}$C and $^{16}$Ne.
When compared to the 3$\alpha$ linear chain configuration in $^{12}$C, at low rotational frequency, the positions of the peripheral clusters in the mirror nuclei
$^{16}$C and $^{16}$Ne appear to be much closer to the central one. The alignment of the valence nucleons from the $\pi$ to the $\sigma$  orbital with increasing rotational frequency, however, reduces the bonding of the outer clusters and their relative position with respect to the central cluster becomes comparable to the one found in $^{12}$C. An additional effect of the alignment of the valence nucleons is an enlargement of the central cluster which, at low rotational frequency, appears to be smaller in size compared to the peripheral ones, but approaches their width at higher rotational frequencies.

The model based on 3D lattice cranking CDFT can be employed in further studies of rotations of cluster structures in $N=Z$ nuclei, or molecular nuclei like the ones considered in the present work. A particularly important point is the possibility to investigate structures that are not constrained by axial symmetry but allow for bending motion. It has been shown that some linear $\alpha$-chain configurations, which also include valence nucleons, are not stable against bending and, therefore, it will be interesting to analyze the effect of collective rotation on such structures.

\begin{acknowledgments}
This work was partly supported by the National Key R\&D Program of China (Contracts No. 2018YFA0404400 and 2017YFE0116700), the National Natural Science Foundation of China (Grants No. 12070131001, 11875075, 11935003, 11975031, and 12141501), the China Postdoctoral Science Foundation under Grant No. 2020M670013, the High-end Foreign Experts Project of Peking University, and the High-performance Computing Platform of Peking University, the QuantiXLie Centre of Excellence, a project co-financed by the Croatian Government and European Union through the European Regional Development Fund - the Competitiveness and Cohesion Operational Programme (KK.01.1.1.01.0004), and the Croatian Science Foundation under the project Uncertainty quantification within the nuclear energy density framework (IP-2018-01-5987).
\end{acknowledgments}

\end{document}